\documentclass[a4paper,12pt]{article}
\title{\LARGE\bf $SLE$ martingales and\\ the Virasoro algebra}
\date{}
\author{}

\def\debut{\begin{eqnarray}}
\def\fin{\end{eqnarray}}
\def\non{\nonumber}
\def\square{ {\hfill \vrule height6pt width6pt depth1pt} }
\usepackage[final]{graphicx}
\begin{document}
\maketitle

\vspace{-1.5 truecm}

\centerline{\large Michel Bauer\footnote[1]{Email:
    bauer@spht.saclay.cea.fr} and Denis Bernard\footnote[2]{Member of
    the CNRS; email: dbernard@spht.saclay.cea.fr}} 

\vspace{.3cm}

\centerline{\large Service de Physique Th\'eorique de Saclay}
\centerline{CEA/DSM/SPhT, Unit\'e de recherche associ\'ee au CNRS}
\centerline{CEA-Saclay, 91191 Gif-sur-Yvette, France}


\def\mony{Y}
\def\mong{X}

\vspace{1.0 cm}

\begin{abstract}
 We present an explicit relation between representations of the
 Virasoro algebra and polynomial martingales in stochastic Loewner
 evolutions (SLE). We show that the Virasoro algebra is the
 spectrum generating algebra of SLE martingales. This is based on a
 new representation of the Virasoro algebra, inspired by the
 Borel-Weil construction, acting on functions
 depending on coordinates parametrizing conformal maps.

\end{abstract}


\vskip 1.5 truecm


Fractal critical clusters are the cornerstones of criticality,
especially in two dimensions, see eg
refs.\cite{nienhuis,cardy,bertrand}.  Stochastic Loewner evolutions
\cite{schramm0,schramm,LSW} are random processes adapted to a
probabilistic description of such fractals.
The aim of this Letter is to elaborate on the connection between 
stochastic Loewner evolutions (SLE) and conformal field theories (CFT) developed
in ref.\cite{BaBe}. We shall construct new representations of the Virasoro
algebra which allow us to show explicitely that the Virasoro algebra
is the generating algebra of (polynomial) martingales for the SLE
processes. Physically, martingales are observables conserved in mean.
They are essential ingredients for estimating probability of events.
Another approach for connecting SLE to representations of the Virasoro 
algebra has been described in \cite{werner}.

Basic definitions of the stochastic Loewner evolutions and of their
martingales are recalled in the two first sections. The new
representations of the Virasoro algebra we shall construct are
described in sections 3 and 4. They are based on a generalization of
the Borel-Weil construction, which we apply to the Virasoro algebra.
They lead to expressions of the Virasoro generators as first order
differential operators acting on (polynomial) functions depending on 
an infinite set of coordinates parametrizing (germs of) conformal maps.
Although motivated by SLE considerations, this is a result independent
of SLE which may find other applications in CFT, string theories or
connected subjects. The applications to SLE are
presented in sections 5 and 6. In particular, we show
that all polynomial SLE martingales are in the Virasoro orbit obtained
by acting with these Virasoro generators on the constant function. 
Because it deals with polynomial martingales and well defined Virasoro
generators, this construction gives a precise algebraic meaning 
to the statement \cite{BaBe} that CFT gives all SLE martingales.
It is more algebraic but less geometric.

\vskip 1.0 truecm

{\bf 1- SLE basics.}
Stochastic Loewner evolutions are growth processes 
defined via conformal maps which are solutions of Loewner's equation:
$$
\partial_t \mong_t(z)=\frac{2}{\mong_t(z)-\xi_t}\ ,\quad \mong_{t=0}(z)=z
$$
with $\xi_t$ real. The map $\mong_t(z)$ is the
uniformizing map for a simply connected domain ${\bf H_t}$ of the upper half
plane ${\bf H}$, ${\rm Im}z>0$. The map $\mong_t(z)$ is normalized by
$\mong_t(z)=z +2t/z + \cdots$ at infinity. For fixed $z$, it is well-defined up to
the time $\tau_z\leq +\infty$ for which $\mong_{\tau_z}(z)=\xi_{\tau_z}$.  
The sets $K_t=\{z\in{\bf H}:\ \tau_z\leq t\}$ form
an increasing sequence, $K_{t'}\subset K_t$ for $t'<t$, 
and are called the hulls. The
domain ${\bf H_t}$ is ${\bf H}\setminus K_t$.
The $SLE$ processes are defined \cite{schramm0} by choosing 
$\xi_t=\sqrt{\kappa}\, B_t$ with $B_t$ a normalized Brownian
motion and $\kappa$ a real positive parameter so that 
${\bf E}[\xi_t\,\xi_s]=\kappa\,{\rm min}(t,s)$.
In particular, $\dot \xi_t$ are  white-noise variables: 
${\bf E}[\dot \xi_t\, \dot \xi_s]=\kappa\, \delta(t-s)$.
Here and in the following,  
${\bf E}[\cdots]$ denotes expectation value.
It will be convenient to introduce the function 
$\mony_t(z)\equiv \mong_t(z)-\xi_t$ whose It\^o derivative is:
\debut
d \mony_t(z) = \frac{2}{\mony_t(z)}dt - d \xi_t 
\label{loew2}
\fin

The $SLE$ equation (\ref{loew2}) may be turned 
into a hierarchy of differential equations for the
coefficients of the expansion of $\mony_t(z)$ at infinity. Writing 
$\mony_t(z)\equiv \sum_{n\geq 0}a_n z^{1-n}$
with $a_0=1$, $a_1=-\xi_t$, and defining polynomials $p_j$ 
in the variables $a_i$ by $p_1=0$, $p_2=1$ and
$p_j = -\sum_{i=1}^{j-2} a_i\ p_{j-i}$ for $j\geq 3$, so that
$\mony^{-1}_t(z)= \sum_{n=1}^\infty p_nz^{1-n}$,
the Loewner equation (\ref{loew2}) becomes:
\debut
\dot{a}_j=2\ p_j(a_1,\cdots) ,\quad j \geq 2. \label{loewbis}
\fin
 
Since $a_1(t)=-\xi_t$ is a Brownian motion, with continuous
trajectories, eqs.(\ref{loewbis}) form a set of stochastic
differential equations for the $a_j(t)$'s, $j\geq 2$, and the solutions are
continuously differentiable functions of $t$ which vanish at $t=0$ due
to the initial condition $\mony_0(z)=z$. Thus, the Ito differential of any
 (polynomial, say) function $Q(a_1(t),a_2(t),\cdots)$ is
$$\left[-d\xi_t \frac{\partial}{\partial
a_1}+dt\left(\frac{\kappa}{2}\frac{\partial^2}{\partial a_1 \, ^2}+2\sum_{j
  \geq 2}p_j\frac{\partial}{\partial a_j}\right)\right]Q.$$
In particular, $$ \frac{d}{dt}{\bf E}[\,Q(a_1(t),a_2(t),\cdots)\,]=
{\bf E}[\,(\hat A \cdot Q)(a_1(t),a_2(t),\cdots)\,]$$
where the opertator $\hat A$ is the coefficient of $dt$ in the
previous formula : 
\debut
\hat A=\frac{\kappa}{2}\frac{\partial^2}{\partial a_1 \, ^2}+2\sum_{j
  \geq 2}p_j\frac{\partial}{\partial a_j}.
\label{evolsto}
\fin
If one assigns degree $i$ to $a_i$, $\frac{\partial}{\partial a_i}$ is
of degree $-i$ and $p_j$ is homogeneous of degree $j-2$, so that $\hat
A$ is of degree $-2$.

In the following, we shall treat the functions $a_i(t)$ as independant
algebraic indeterminates $a_i$, as already suggested in previous
notations. This requires some justification. We need to show that if
$Q(a_1,a_2,\cdots)$ is a nonzero abstract polynomial, the function
$Q(a_1(t),a_2(t),\cdots)$ cannot vanish for every realization of
$\xi_t$ and every $t$.  Indeed, assume the countrary and take a
counterexample $Q(a_1,a_2,\cdots)$ of minimal degree. As
$Q(a_1(t),a_2(t),\cdots)\equiv 0$, the Ito differential of $Q$
vanishes identically as well : $\left[-d\xi_t \frac{\partial}{\partial
    a_1}+dt{\hat A}\right]Q(a_1(t),a_2(t),\cdots)\equiv 0$.
Multiplication by $d\xi_t$ yields $-dt\frac{\partial Q}{\partial
  a_1}(a_1(t),a_2(t),\cdots)\equiv 0$ which can be plugged back into
the original equation. So ${\hat A}Q(a_1(t),a_2(t),\cdots)\equiv 0$ 
and $\frac{\partial Q}{\partial a_1}(a_1(t),a_2(t),\cdots)\equiv 0$.
Because ${\hat A}$ and $\frac{\partial }{\partial a_1}$ decrease the
degree and $Q$ is a counterexample of minimal degree, ${\hat A}Q$ and
$\frac{\partial Q}{\partial a_1}$ vanish as abstract polynomials. So the
question whether the functions $a_i(t)$ are algebraically independant
is reduced to the purely algebraic question whether the system of
linear algebraic equations $\frac{\partial Q}{\partial a_1}={\hat
  A}Q=0$ has only the constant polynomials as solutions. 
This will be proved at the end of section 4.

\vskip 1.0 truecm

{\bf 2- SLE martingales.}
The set ${\cal F}$ of polynomial functions in the $a_j$ forms a graded
vector space ${\cal F}\equiv \oplus_{n\geq 0} {\cal F}_n$, with elements of
${\cal F}_n$ homogeneous polynomials of degree $n$. The operator $\hat
A$ maps ${\cal F}_{n+2}$ into ${\cal F}_{n}$.

Polynomial martingales are, by definition, polynomials in the $a_j$
annihilated by $\hat A$. Their set ${\cal M}={\rm ker}\hat A$ is
graded: ${\cal M}\equiv \oplus_{n \geq 0} {\cal M}_n$ with ${\cal
M}_n\subset {\cal F}_n$.  The low degree martingales are: 
\debut {\cal
M}_1 &:& a_1 \non\\ 
{\cal M}_2 &:& 2a_1^2-\kappa a_2 \non\\ 
{\cal M}_3 &:& 2a_1^3-3\kappa a_1a_2,\ a_3 + a_1a_2\non 
\fin 
Let ${\rm
char}\, {\cal M}=\sum_{n\geq 0}q^n{\rm dim}\, {\cal M}_n$ with ${\cal
M}_n={\rm ker}\hat A\vert_{{\cal F}_n}$.  Crucial to the sequel is
the character formula  
\debut
{\rm char}\;{\cal M}=\frac{1-q^2}{\prod_{j\geq 0}(1-q^j)}
\label{charac}
\fin 
This may be compared to 
${\rm char}\;{\cal F}=\prod_{j\geq 0}(1-q^j)^{-1}$, 
in particular ${\rm dim}\;{\cal M}_n<{\rm dim}\;{\cal F}_n$ for $n\geq 
2$. 

We shall give a proof of eq.(\ref{charac}) using the general machinery
in section 5.  A direct argument can be organized as follows. We want
to show that the sequence ${\cal M}_{n+2} \stackrel{\hat
  A}{\rightarrow} {\cal M}_{n} \rightarrow 0$ is exact, i.e. that
$\hat A ({\cal M}_{n+2})={\cal M}_{n}$ for $n\geq 0$. Decompose $\hat
A= {\hat A}'+ {\hat A}''$ where ${\hat A}'\equiv 2
\frac{\partial}{\partial a_2}$. It is clear that ${\hat A}' ({\cal
  M}_{n+2})={\cal M}_{n}$ for $n\geq 0$ because if $Q(a_1,a_2,\cdots)$
is any polynomial in ${\cal M}_{n}$, we can set ${\hat
  I}Q(a_1,a_2,\cdots)\equiv \frac{1}{2}\int_0^{a_2}da\,
Q(a_1,a,a_3\cdots)$, which satisfies clearly $ {\hat A}'{\hat I}Q=Q$.
Now we do perturbation theory.  Starting from $Q\neq 0$, we define two
sequences $q_n, n\geq 0$ and $r_n,n\geq 1$ by $q_0\equiv Q$,
$r_1\equiv {\hat I} q_0$, $q_1\equiv q_0-{\hat A}r_1=-{\hat A}''{\hat
  I}q_0$, $r_2\equiv {\hat I}q_1$, $q_2\equiv q_1-{\hat A}r_2=-{\hat
  A}''{\hat I}q_1$, $\cdots$. The key point if that these sequences
stop. Indeed, if $q_k$ is non zero, then its total degree is that of
$q_{k-1}$ but its degree in $a_2$ is at least one more than that of
$q_{k-1}$ : ${\hat I}$ increases the degree in the variable $a_2$ by
one unit and ${\hat A}''$ contains no derivative with respect to
$a_2$. So $q_k$ and then $r_{k+1}$ have to vanish for large enough
$k$. Hence one can sum the definition ${\hat A}r_{k+1}=q_k-q_{k+1}$
over $k$, leading to a telescopic cancellation ${\hat A}\sum_k
r_k=q_0$, showing that ${\hat A}$ is onto.

\vskip 1.0 truecm

{\bf 3- Group theoretical background.} 
Let us recall a few basic facts concerning the Borel-Weil
construction in group theory. Consider for instance a simply
connected compact Lie group $G$. 
The group acts on itself by left or right
multiplication. This induces a representation of the Lie algebra ${\rm
  Lie}G$ on functions on $G$ by left or right invariant vector fields:
\debut
(X\cdot\nabla^l) f(g)&=& \frac{d}{du}f(ge^{uX})\vert_{u=0},\non\\
(X\cdot\nabla^r) f(g)&=& \frac{d}{du}f(e^{uX}g)\vert_{u=0}\non
\fin
for any $X\in {\rm Lie} G$. They form a representation of ${\rm Lie}
G$ since $[X\cdot\nabla^l , Y\cdot\nabla^l]=[X,Y]\cdot\nabla^l$
and $[X\cdot\nabla^r , Y\cdot\nabla^r]=-[X,Y]\cdot\nabla^r$.

Let us choose a Cartan subgroup $H$, and let $N_\pm$ be the
associated nilpotent subgroups and $B_\pm=HN_\pm$ be 
the corresponding Borel subgroups
\footnote{For $G=SU(N)$, $H$ is made of diagonal complex matrices and
$N_\pm$ of lower (upper) triangular matrices with $1$ along the
diagonal.}.
At least in a neighbourhood of the identity, elements $g$ of $G$
may be factorized according to the Gaussian decomposition as
$g=n_+hn_-$ with $h\in H$ and $n_\pm\in N_\pm$. We set
$g_+=n_+h$, $g_0=h$ and $g_-=n_-$, the components of $g$ in $B_+$,
in $H$ and in $N_-$, respectively. For elements $X\in {\rm Lie}G$, we
shall denote by $X_+$, $X_0$ and $X_-$ their components in ${\rm
  Lie}B_+$, in ${\rm Lie} H$ and in ${\rm Lie}N_-$, respectively.

One may define two actions of $G$ on $N_-$ by:
\debut
 l_x(g)&\equiv& (gxg^{-1})_+\,^{-1}\,g\, x= (gxg^{-1})_-\, g \non\\
 r_x(g) &\equiv& x^{-1}\, g\, (g^{-1}xg)_+ = g\, (g^{-1}xg)_-\,^{-1} \non
\fin
for $g\in N_-$ and $x\in g$. They act on $N_-$ since $l_x(g)\in N_-$
and $r_x(g)\in N_-$ for $g\in N_-$. They form anti-representations of $G$:
$$ l_y( l_x(g) )= l_{xy}(g),\quad r_y( r_x(g) )= r_{xy}(g)$$
because
$(g xy g^{-1})_+= (gxg^{-1})_+\, (l_x(g)yl_x(g)^{-1})_+ $.

The Borel-Weil construction consists in defining an action of the group
$G$ on sections of line bundles over the quotient space $B_+\backslash G^{\bf C}$. 
Sections of $B_+\backslash G^{\bf C}$ may be viewed as functions $S^l(g)$ 
on $G^{\bf C}$ such that
\debut
S^l(n_+g)&=& S^l(g),\quad ~~~~~~ n_+\in N_+ \non\\
S^l(hg) &=& \chi(h)\, S^l(g),\quad h\in H \label{section}
\fin 
with $\chi(h)$ a ${\bf C}$-valued $H$-character such that
$\chi(h_1h_2)=\chi(h_1)\chi(h_2)$. Such a character is specified by 
a weight $\omega\in ({\rm Lie}H)^*$ via $\chi(e^X)=\exp{-(\omega,X)}$
for $X\in {\rm Lie}H$.

The group $G$ acts on such sections by right
multiplication: $(L_x\cdot S^l)(g)\equiv S^l(gx)$ for $x\in G$. 
This action defines a representation of $G$:
$L_y\cdot(L_x\cdot S^l)(g)=(L_{yx}\cdot S^l)(g)$.
Since $B_+\backslash G^{\bf C}$ may locally be identified with $N_-$, we may
choose a gauge in which $g\in N_-$ and view $S^l$ as functions on
$N_-$ with specific transformation properties.  The action of $G$ then
reads: 
\debut 
(L_x\cdot S^l)(g)= \chi((gxg^{-1})_0)\,
S^l(l_x(g)), \quad x\in G,\ g\in N_-
\label{BWaction1}
\fin
Infinitesimally, this action may be presented as first order
differential operator:
\debut
(D^l_X\cdot S^l)(g) &=& 
\frac{d}{du} (L_{e^{uX}}\cdot S^l)(g)\vert_{u=0}
\label{BWaction2}\\
&=& \Big((X\cdot\nabla^l) - ((gXg^{-1})_+\cdot\nabla^r)\Big)S^l(g) 
-(\omega, (gXg^{-1})_0)S^l(g) \non
\fin
with $\nabla^l$ and $\nabla^r$ the left and right invariant vector fields.
By construction $[D^l_X,D^l_Y]=D^l_{[X,Y]}$. Note that $D_X^l$ coincides
with $X\cdot \nabla^l$ for $X\in {\rm Lie} N_-$.

A similar construction applies to the right quotient $G^{\bf C}/B_+$ and its
sections $S^r(g)$, defined similarly as in eq.(\ref{section}) but with
right instead of left multiplications. The group $G$ acts on sections
$S^r$ by left multiplications: $(R_x\cdot S^r)(g)=S^r(x^{-1}g)$.
In the gauge in which $g\in N_-$, this action reads:
\debut
(R_x\cdot S^r)(g)\equiv \chi((g^{-1}xg)_0)\,
S^r(r_x(g)), \quad x\in G,\ g\in N_-
\label{BWaction3}
\fin
Infinitesimally:
\debut
(D^r_X\cdot S^l)(g) = \Big(((g^{-1}Xg)_+\cdot\nabla^l) -
(X\cdot\nabla^r)\Big)S^r(g) - (\omega, (g^{-1}Xg)_0)S^l(g)
\label{BWaction4}
\fin
By construction $[D^r_X,D^r_Y]=D^r_{[X,Y]}$.

These left and right actions are linked by
the relation $r_x(g)l_x(g^{-1})=1$.
Therefore, if $D_X^l$ are represented as first order differential
operators in  a specific set of coordinates parametrizing $g\in N_-$,
then $D_X^r$ will be represented by the same differential operators but
in the coordinates parametrizing the inverse element $g^{-1}$.

\vskip 1.0 truecm

{\bf 4- Differential representations of the Virasoro algebra.} 
We now apply the previous construction to the Virasoro algebra
with generators $L_n$, $n$ integers, and relations:
$$ 
[L_n,L_m]= (n-m)L_{n+m} +\frac{\hat c}{12}n(n^2-1)\delta_{n+m,0}
$$
with $\hat c$ central. By convention, ${\rm Lie}N_-$ is generated by
the $L_n$ with $n<0$, ${\rm Lie}H$ by $L_0$ and $\hat c$, and ${\rm
  Lie}N_+$ by the $L_n$ with $n>0$.  

We have to select a set of coordinates in $N_-$, at least in a
neighbourhood of the identity. This is provided by
looking at the representation in which the Virasoro
generators are represented by $\ell_n=-z^{n+1}\partial_z$. 
The $N_-$ orbits of a point $z$ in the complex plane define (germs
of) complex maps $w(z)$ with a simple pole at infinity:
\debut
w(z)\equiv g\, z\, g^{-1} = \sum_{n\geq 0} a_n\, z^{1-n},\quad g\in
N_- 
\label{coordi}
\fin
with $a_0=1$. The $a_n$ form a set of coordinates parametrizing
elements of $N_-$.
We shall also need the inverse map $z(w)$:
\debut
z(w) = \sum_{n\geq 0} b_n w^{1-n} \label{zdew}
\fin
The $b_n$ are polynomials in the $a_j$ of degree $n$ and $b_0=1$.

Let us first deal with sections of the left quotient 
$B_+\backslash G^{\bf  C}$. To define them we have to specify the
$H$--character, or equivalently the weight $\omega\in ({\rm Lie}H)^*$. 
It is specified by 
two numbers $\delta$ and $c$ such that $(\omega,L_0)=\delta$ and 
$(\omega,\hat c)=c$. The action of the Virasoro generators on
functions of the $a_j$ is then defined by the formula
(\ref{BWaction2}). One may view $w(z)$ as functions of the $a_j$ and
use it as generating functions. We then have\footnote{Here, we use the
convention that for a Laurent series $h(z)=\sum_j h_jz^j$,
$(h(z))_0=h_0$ and $(h(z))_+=\sum_{j\geq 1} h_j z^j$.}:
\vskip 0.5 truecm

{\bf Proposition.}\\
{\it i) The action of the Virasoro algebra on
$w(z)=\sum_{n\geq 0}a_n z^{1-n}$ specified by eq.(\ref{BWaction2}) reads:
\debut
D_n^l\cdot w(z) &=& -w(z)^{n+1} 
+ \Big( w^{n+1}\frac{dz}{dw} \Big)_+ (\frac{dw}{dz}) \label{bigbig}\\
& & ~~~~~ -\Big(\, \delta\, (gL_ng^{-1})\vert_{L_0} 
+ c\, (gL_ng^{-1})\vert_{\hat c} \,\Big) w(z) \non
\fin
with
$$ (gL_ng^{-1})\vert_{L_0}=\Big( z^{-1}w^{n+1}\,\frac{dz}{dw}\Big)_0,\quad
(gL_ng^{-1})\vert_{\hat c}= -\frac{1}{12} 
\Big(zw^{n+1}\, S_z(w)\,\frac{dz}{dw}\Big)_0
$$
where $S_z(w)=\frac{w'''}{w'}-\frac{3}{2}(\frac{w''}{w'})^2$ is the
Schwarzian derivative of $w(z)$ with respect to $z$.\\
ii) The first order differential operators $D_n^l$ in the $a_j$
satisfy the Virasoro algebra 
$$[D_n^l,D_m^l]=(n-m)D^l_{n+m}-\frac{c}{12}n(n^2-1)\delta_{n+m,0}$$
and are such that $D_n^l\cdot 1=0$ for $n<0$ and $D_0^l\cdot 1 = -\delta$.}

{\bf Proof.} The three terms in eq.(\ref{bigbig}) correspond to the
three terms in eq.(\ref{BWaction2}). They may be computed one by one
using the following relations, which we just quote without proofs. 
First one has:
\debut 
(L_n.\nabla^l)w(z)&=& g[\ell_n,z]g^{-1}=-w^{n+1}\non\\
(L_n.\nabla^r)w(z)&=& [\ell_n, gzg^{-1}]=-z^{n+1} (\frac{dw}{dz})\non
\fin
Then, $gL_ng^{-1}$ is evaluated using the transformation properties
\cite{bpz} of the stress tensor $T(z)=\sum_n L_n z^{-n-2}$:
$$g T(w) g^{-1}\, dw^2= T(z(w))\, dz^2 - \frac{\hat c}{12}
S_z(w)\, dz^2$$ 
Recall that $S_w(z)dw^2=-S_z(w)dz^2$. This gives:
$$ gL_ng^{-1}=\sum_{s\leq n} L_s\
\Big(\frac{w^{n+1}}{z^{s+1}}\frac{dz}{dw}\Big)_0
-\frac{\hat c}{12} \Big(zw^{n+1}\, S_z(w)\,\frac{dz}{dw}\Big)_0$$
In particular, $(gL_1g^{-1})\vert_{L_0}=2a_1$,
$(gL_2g^{-1})\vert_{L_0}=3a_1^2+4a_2$ and $(gL_2g^{-1})\vert_{\hat
  c}=a_2/2$. 
\square
\vskip 0.5 truecm

The differential operators $D_n^l$ may be written explicitely.
They are of the form $D_n^l={\cal D}_n^l(a)-\delta
d_\delta^{(n)}-cd_c^{(n)}$ with ${\cal D}_n^l(a)=\sum_j
q_j^{(n)}\partial_{a_{j-n}}$, and $d_\delta^{(n)}$,
$d_c^{(n)}$ and $q_j^{(n)}$ homogeneous polynomials.
The first few are:
\debut
D_{-2}^l&=& {\cal D}_{-2}^l(a) \non\\
D_{-1}^l&=& {\cal D}_{-1}^l(a) \label{diffvir}\\
D_{0}^l&=& {\cal D}_{0}^l(a) -\delta \non\\
D_{1}^l&=& {\cal D}_{1}^l(a) -2\delta a_1 \non\\
D_{2}^l&=& {\cal D}_{2}^l(a)-(3a_1^2+ 4a_2)\delta - \frac{c}{2}a_2 \non
\fin
with ${\cal D}_{n}^l(a)$ vector fields given by:
\debut
{\cal D}^l_{-2}(a)&=& -\sum_{j\geq 2} p_{j}(a)\frac{\partial}{\partial a_{j}}\non\\ 
{\cal D}^l_{-1}(a)&=& -\frac{\partial}{\partial a_1} \non\\
{\cal D}^l_{0}(a)&=& -\sum_{j\geq 1} ja_j\frac{\partial}{\partial a_j}\non\\
{\cal D}^l_{1}(a)&=& -\sum_{j\geq 1}\Big( \sum_{p+q=j+1} a_pa_q +
ja_{j+1} + 2(j-1) a_1a_j \Big) \frac{\partial}{\partial a_j}\non\\
{\cal D}^l_{2}(a)&=& -\sum_{j\geq 1} \Big( \sum_{p+q+r=j+2} a_pa_qa_r +
(j+1)a_{j+2}  \non\\
& & ~~~~~~~~~~~~  + 3j a_1a_{j+1}+ (j-1)(3a_1^2+4a_2)a_j \Big) 
\frac{\partial}{\partial a_j}\non
\fin
The above sums include the terms with $a_0=1$.
These five operators generate the whole Virasoro algebra.
They act on polynomial functions of the $a_j$.

Note that it was imperative to consider sections of $B_+\backslash
G^{\bf C}$ associated to non trivial $H$--characters in order to get
representations of the Virasoro algebra with non vanishing central
charges.  To make contact with usual highest weight representations,
one may define generators $\tilde L_n\equiv -D_{-n}$ which satisfy the
Virasoro algebra with central charge $c$. They are such that $\tilde
L_n\cdot 1=0$ for $n>0$ and $\tilde L_0\cdot 1=\delta$, so that $1$ is
a highest weight vector for the $\tilde L_n$.

As explained above, one may define another action of the Virasoro
algebra using the right quotient $G^{\bf C}/B_+$.  Its generators
$D_n^r$ are defined via eq.(\ref{BWaction4}). 
According to the last remark of the previous section, one goes from
$D_n^l$ to $D_n^r$ by exchanging the role played by $g$ and its
inverse, so that $D_n^r$ coincides with $D_n^l$ but with the variables
$a_j$ replaced by those parametrizing the inverse map:
\debut
 D_n^r = D_n^l(a_j\to b_j) \label{Dright}
\fin
The two representations $D_n^r$ and $D_n^l$ do not commmute. 

Let us remark that although the space of (polynomial) functions in the $a_j$ may
be identified with a Fock space, these representations are not the usual
free field representations used in conformal field theory \cite{FF}. 
As far as we know, these representations
were not previously described in the literature.
They are similar in spirit to the representations of affine Kac-Moody
algebras studied in \cite{BerFelder}.

We end this section by showing that a polynomial $Q$ such that
$\frac{\partial Q}{\partial a_1}=0$ and $\hat A\cdot Q=0$ is constant,
thereby completing the proof, started below eq.(\ref{evolsto}),
that the functions $a_n(t)$ are
algebraically independent. The key observation is that 
$D^l_{-n}$ has the form $D^l_{-n}=-\frac{\partial}{\partial
  a_n}-\sum_{m\geq n+1}p_{n,m}\frac{\partial}{\partial
  a_m}$, where $p_{n,m}$ is a polynomial,
$p_{1,m}=0,p_{2,m}=p_m,\cdots$. This results from  the recursive
definition of $D_{-n}^l$ and the fact that, as a polynomial in $a_1$,
 $p_n=-(-a_1)^{n-2}+$ terms of lower degree.
By hypothesis, the polynomial $Q$ is
annihilated by $\frac{\partial}{\partial a_1}=-D^l_{-1}$ and 
$\hat A=\frac{\kappa}{2}(D^l_{-1})^2-2D^l_{-2}$. Hence it is annihilated by all
$D^l_{-n}$'s. The polynomial $Q$ depends effectively on only a finite
number of variables : there is a minimal $n$ such that $\frac{\partial
  Q}{\partial a_m}=0$, $m>n$. If $n>0$,  $D^l_{-n}Q=0$ implies that
$\frac{\partial Q}{\partial a_n}=0$ contradicting minimality. So $n=0$
and $Q$ is constant, as was to be proved.

\vskip 1.0 truecm

{\bf 5- Martingale generating algebra.} 
Let us now make contact between the stochastic Loewner equation 
and the representations of the Virasoro algebra we just define.

Comparing the evolution operator $\hat A$, eq.(\ref{evolsto}),
and the operator $D_n^l$, eqs.(\ref{diffvir}), it is clear that 
one has the following identification:
$$ 
\hat A = \frac{\kappa}{2} (D^l_{-1}\,) ^2 - 2 D^l_{-2}
$$
In other words, the stochastic evolution (\ref{evolsto}) is associated
with the action the Virasoro algebra on sections of the left quotient
$B_+\backslash G^{\bf C}$. 

On the other hand, the martingale generating algebra is not
constructed using the representation $D_n^l$, since it does not
commute with $\hat A$, but using the representation $D_n^r$ based on
the right quotient $G^{\bf C}/B_+$. Indeed, we have:
\vskip 0.5 truecm

{\bf Proposition.}
{\it For $c=\frac{(3\kappa-8)(6-\kappa)}{2\kappa}$ and
$\delta=\frac{6-\kappa}{2\kappa}$, one has the commutation relations:
\debut
\Big[\, \hat A\, ,\, D_n^r\, \Big] = \widehat q_n(a_1,\cdots)\, \hat A
\label{martincom}
\fin 
with $\widehat q_n(a_1,\cdots)$ homogeneous polynomials in the
$a_j$ of degree $n$.\\ In particular, the generators $D_n^r$ act on
${\rm ker}\hat A$, so that if $f\in {\rm ker}\hat A$ then $D_n^r\cdot
f\in {\rm ker}\hat A$.}

{\bf Proof.} This follows by construction.
For $n<0$, $D_n^r$ coincide with the right invariant vector fields
which commute with the left invariant vector fields and thus with
$\hat A$. For $n\geq 0$, eq.(\ref{martincom}) may  be checked directly
using the following relation 
\debut
[D_Y^l,D_X^r]\cdot f(g) &=& \Big( [Y, (g^{-1}Xg)_+]-[Y,
(g^{-1}Xg)]_+\Big)\cdot \nabla^l f(g) \non\\
&& +(\omega,[Y,(g^{-1}Xg)]_0)\, f(g)\non
\fin
valid for $Y\in {\rm Lie}N_-$ and $X\in {\rm Lie}G$.
For instance, applying this formula for $X=L_{1}$ leads to :
$$ [\hat A, D_1^r]\cdot f = (6 -\kappa(2\delta
+1))(L_{-1}\cdot\nabla^l)\cdot f - 4 b_1 (\hat A\cdot f) $$
The first term in the right hand side vanishes for $2\kappa \delta =
6-\kappa$. Similarly, $X=L_2$ gives:
\debut
[\hat A, D_2^r]\cdot f &=& 
3(6-\kappa - 2\kappa \delta)b_1(L_{-1}\cdot\nabla^l)\cdot f\non\\
&& +((8-3\kappa)\delta + c)f -2(3b_1^2 + 4b_2)(\hat A\cdot f)\non
\fin
The first two terms in the r.h.s. vanish for $2\kappa \delta
=6-\kappa$ and $c=(3\kappa-8)\delta$. Note that $\widehat q_1=-4b_1$
and $\widehat q_2=-2(3b_1^2+4b_2)$. The higher degree polynomials
$\widehat q_n$ are recursively determined by 
$(n-m)\widehat q_{n+m}=[D_n^r,\widehat q_m]-[D_m^r,\widehat q_n]$.
\square
\vskip 0.5 truecm

Let us first prove eq.(\ref{charac}) using this proposition.  Since
${\cal M}_n={\rm ker}\hat A\vert_{{\cal F}_n}$ and since $\hat A$ has
degree $-2$, proving formula (\ref{charac}) amounts to show that $\hat
A:{\cal F}_{n+2}\to {\cal F}_{n}$ is surjective. We do it by recursion
using the fact that $D_{-j}^r$, $j\geq 1$, commute with $\hat A$. Let
$u\in {\cal F}_{n+2}$.  By the recursion hypothesis, there exist
$w_1\in {\cal F}_{n-1}$ and $w_2\in {\cal F}_{n-2}$ such that
$D_{-1}^ru=\hat A w_1$ and $D_{-2}^ru=\hat A w_2$. Let us define
recursively $w_j\in {\cal F}_{n-j}$ by
$(j-1)w_{j+1}=D_{-1}^rw_j-D_{-j}^rw_1$. By construction they satisfy:
i) $D_{-j}^ru=\hat A w_j$ and ii)
$D_{-j}^rw_i-D_{-i}^rw_j=(i-j)w_{i+j}$. Relation ii) is the
integrability condition for the existence of $v\in {\cal F}_n$ such
that $w_j=D_{-j}^rv$. Relation i) then gives $D_{-j}^ru = D_{-j}^r
\hat A v$ for all $j\geq 1$. This implies $u= \hat A v$, meaning that
$\hat A$ is surjective. Note that this proof is dual to the proof
given at the end of section 4 that the functions $a_n(t)$ are
algebraically independant.

Let us now remark that acting successively with $D_n^r$ on the
constant function $1$ generates $SLE$ martingales. For instance: 
\debut
D_1^r\cdot 1 &=& (\frac{6-\kappa}{\kappa})\, a_1 \non\\
4D_2^r \cdot 1 = -\kappa\, D_1^r\, ^2 \cdot 1 &=&
3(\frac{6-\kappa}{\kappa})\,(\kappa a_2 -2 a_1^2)\non 
\fin 
Recall that the operators $D_n^r$ are obtained from the $D_n^l$,
eq.(\ref{diffvir}), by replacing $a_j$ by $b_j$.  
More generally, the space ${\cal P}$
of polynomials in the $a_j$ generated by successive actions of the
$D_n^r$ on the constant function, that is
$${\cal P}= {\rm vect.}<\prod_jD_{n_j}^r\cdot 1>,$$
is made of martingales and so it is embedded in
${\cal M}$. By construction it carries a representation of the
Virasoro algebra. It is well know \cite{FF} that its character is
$$ {\rm char}{\cal P}= \frac{1-q^2}{\prod_{j\geq 0}(1-q^j)}$$
So ${\rm char}{\cal P}={\rm char}{\cal M}$ and  
${\cal P}\equiv{\cal M}$. 

In other words, all polynomial $SLE$
martingales are generated by successive actions of the Virasoro
differential operators associated to the right quotient.

\vskip 1.0 truecm

{\bf 6- Lifted SLE.} Let us finally make contact with the group
theoretical formulation of the stochastic Loewner evolution proposed
in ref.\cite{BaBe}. There, the $SLE$ was lifted to a Markov process
in the nilpotent subgroup $N_-$ of the Virasoro group defined by:
$$ dg_t = g_t (-2L_{-2}\, dt + L_{-1}\, d\xi_t), \quad g_{t=0}=1.$$ 
The associated stochastic evolution operator, acting on function of
$g_t$, was identified with
$$ A = -2 (L_{-2}\cdot \nabla^l) + \frac{\kappa}{2} (L_{-1}\cdot
\nabla^l)^2 $$
See ref.\cite{BaBe} for details. The random group element $g_t$ is
related to the random conformal map $\mony_t$ by $\mony_t(z)=g_tzg^{-1}_t$ 
in the representation with $\ell_n=-z^{n+1}\partial_z$.
Since $D_n^l$ is simply $(L_n\cdot\nabla^l)$ for $n<0$,
the operators $A$ and $\hat A$ clearly coincide.

A generating function of $SLE$ martingales was
identified in \cite{BaBe}. It is given by the vector 
$g_t|\omega_\kappa\rangle$ which takes values in the irreducible (for
generic $\kappa$) Virasoro module, called ${\cal H}_{1,2}$, 
with highest weight vector $|\omega_\kappa\rangle$
of central charge $c=\frac{(3\kappa-8)(6-\kappa)}{2\kappa}$ and
conformal weight $\delta=\frac{6-\kappa}{2\kappa}$.
In a graded basis of ${\cal H}_{1,2}$,
the components of $g_t|\omega_\kappa\rangle$
are polynomial $SLE$ martingales by construction. As is well known,
 ${\rm char}{\cal H}_{1,2}={\rm char}{\cal P}$ 
and this allows us to identifies ${\cal H}_{1,2}$
with ${\cal P}$, so that the  $SLE$ martingales generated by
$g_t|\omega_\kappa\rangle$ coincide with those
obtained above by successive actions with the $D_n^r$.
This actually follows by construction, since in the Borel-Weil
construction, sections $S^r(g)$ of $G^{\bf C}/B_+$ may be identified with
matrix element $S^r(g)=\langle\nu|g|\omega\rangle$ with
$|\omega\rangle$ highest weight vector.

\vskip 1.0 truecm

{\bf Acknowledgement:} Work supported in part by EC contract number
HPRN-CT-2002-00325 of the EUCLID research training network.


\end{document}